**4**

*Altmetrics are tools for measuring the impact of research beyond scientific communities. In general, they measure online mentions of scholarly outputs, such as on online social networks, blogs, and news sites. Some stakeholders in higher education have championed altmetrics as a new way to understand research impact and as an alternative or supplement to bibliometrics. Contrastingly, others have criticized altmetrics for being ill conceived and limited in their use. This chapter explores the values and limits of altmetrics, including their role in evaluating, promoting, and disseminating research.*

# The Values and Limits of Altmetrics

*Grischa Fraumann*

Alternative metrics, or "altmetrics", are tools for measuring online mentions of scholarly outputs, such as mentions on online social networks, blogs, news sites, and Wikipedia. Altmetrics track and count the mentions of scholarly outputs on social media, news sites, policy sites, and social bookmarking sites, and aggregate the number of mentions. This allows observers to see how many times Internet users have viewed, discussed, followed, shared, adapted, and/or downloaded a research study. One definition of altmetrics is as follows:

> Altmetrics are non-traditional metrics that cover not just citation counts but also downloads, social media shares, and other measures of impact of research outputs. The term is variously used to mean "alternative metrics" or "article level metrics", and it encompasses webometrics, or cybermetrics, which measure the features and relationships of online items, such as websites and log files. The rise of new social media has created an additional stream of work under the label altmetrics. These are indicators derived from [online social networks], such as Twitter, [and] Mendeley . . . with data gathered automati-cally by computer programs. (Wilsdon et al., 2015, p. 5)

Compared to the traditional method of counting citations, this approach provides some advantages, such as fast recognition of mentions of scholarly papers online. Nevertheless, it is important to note that these mentions do not necessarily correlate with the quality of a scholarly output; they mainly visualize the amount of online attention. Most scholars agree that online visibility has become a necessity for higher education institutions (HEIs) and it is important to share posts on social media to connect with stakeholders, such as research funders, policy makers, and the wider public. This chapter provides an overview of the values and limits of altmetrics, and it describes the role altmetrics might play within institutional research (IR).

**Background**

The concept of "altmetrics" emerged in 2010, and it was postulated in the altmetrics manifesto by Priem, Taraborelli, Groth, & Neylon, 2010 *Altmetrics: A manifesto* (Howard, 2013). The concept has undergone rapid development in academia over the last few years, and altmetrics has gained increasing attention through several policy initiatives, such as the ones by the European Commission and STAR METRICS in the United States (STAR METRICS, 2017).

Companies external to HEIs, or so-called data aggregators, provide altmetrics. Altmetric.com, one of the largest altmetrics data aggregators had curated "over 10 million research outputs" in the Altmetric Explorer as of June 6, 2017 (Altmetric.com, 2017). PlumX Altmetrics Dashboard is a system similar to the Altmetric Explorer and is provided by the company Plum Analytics. The PlumX Dashboard is an online system used to visualize the societal impact of institutional members (e.g., a HEI) in altmetrics sources and bibliometric databases. Plum Analytics had covered 52.6 million research outputs as of June 7, 2017 (Plum Analytics, 2017). Furthermore, Altmetric.com includes citation counts from Elsevier's Scopus database while Plum Analytics includes citations from Clarivate Analytics' Web of Science. Studies of altmetrics continue to be published each year, and altmetrics are already being called an established research field (Gauch & Blümel, 2016; Robinson-García, Torres-Salinas, Zahedi, & Costas, 2014). Moreover, major intergovernmental organizations, such as the Organisation for Economic Co-operation and Development discuss the use of altmetrics (OECD, 2016).

Research impact is closely related to the concept of altmetrics. As the OECD (2016, p. 143) stated in the *Science, Technology and Innovation Outlook 2016*, "Altmetrics … are likely to be increasingly used alongside more traditional bibliometrics to assess research impacts." This line of thought implies that one might relate these online mentions to a kind of impact research has on the wider public or society outside the scientific community because anyone with an Internet connection should be able to engage with (open access) scholarly outputs online, even though only a fraction of such users may actually do so. Altmetrics data aggregators weigh sources differently; for example, a mention of a research output on a news site increases the altmetrics score more than does a mention on Twitter or Facebook (Altmetric.com, n.d.). However, altmetrics data aggregators do not "rank" online users that have mentioned a scholarly study, it just provides counts of mentions, views, or interactions. Further development of the uses of altmetrics is well-documented (Bornmann, 2014; CWTS, 2017; Holmberg, 2016; Liu & Adie, 2013; Piwowar, 2013; Priem et al., 2010; Robinson-García et al., 2014; Thelwall, Haustein, Larivière, Sugimoto, & Bornmann, 2013) with perhaps the most clear example being the possibility of it as a method for introducing Web mentions into researchers' biographies (see Aaltojarvi, Arminen, Auranen, & Pasanen, 2008).

**The Role of Altmetrics in the Higher Education Sector**

Altmetrics are closely related to another phenomenon, that is, open science (e.g., open access publishing). In this respect, altmetrics can provide evidence of use of open access publishing and open access to research data as they record and recognize each individual interaction with such research outputs. The open science movement has frequently advocated this kind of measurement (S. Niinimäki, personal communication, September 19, 2016; Fecher & Friesike, 2014). Several HEIs have implemented altmetrics tools displaying the altmetrics counts of research outputs on institutional repositories. Furthermore, journals, publisher websites, and large information systems, such as Scientific Electronic Library Online (SciELO), the largest open access repository in Latin America, South Africa, and Spain, display altmetrics as well (Packer, Cop, Luccisano, Ramalho, & Spinak, 2014). SciELO uses Altmetric.com as a data aggregators, and the publications and their altmetrics counts are available on ScienceOpen, a large open science platform provided by a private company. These are attempts to show the interactions among research outputs, researchers, and the general public.

   Altmetrics are playing an increasingly important role in debates in higher education concerning accountability, evaluation, and performance of HEIs, as well as scholarly communication (Adie & Roe, 2013; Alhoori & Furuta, 2014; Bar-Ilan et al., 2012; Leibniz Gemeinschaft, n.d.; Mounce, 2013; van Noorden, 2014). Scholarly communication in particular has undergone a significant shift into the Internet and social media in recent years. In short, altmetrics have become part of the debate on the impact or value created by scholarly research (Auranen, 2006; Bornmann, 2012, 2014; Bornmann & Marx, 2014; Kohtamäki, 2011; Meijer, 2012; Wallace & Ràfols, 2015). This has occurred in large part as a response to demands for the exact measurement of research impact issued by external stakeholders, particularly research funders (National Information Standards Organization, 2016; Sarli, Dubinsky, & Holmes, 2010; STAR METRICS, 2017; van Noorden, 2014). Despite the many challenges and shortcomings of altmetrics, some stakeholders, such as research funders, have argued that in the future, these tools could partly answer the question of return on investment in research, because mentions of studies outside the scientific community can suggest evidence of a societal impact.

To illustrate, several HEIs in the United States and globally use the institutional platform created by Altmetric.com; among these HEIs are also research funding organizations such as the British Wellcome Trust (Thelwall, Kousha, Dinsmore, & Dolby, 2016; Wellcome Trust, 2014). According to Altmetric.com, some researchers are now including their "altmetric attention score" on the CVs they attach to funding proposals (Chimes, 2014). However, it is unclear how widespread the use of altmetrics is in the higher education sector, because marketing materials mention only a few selected examples, but there are no user statistics available.

Considering these developments, it is essential to ensure unmanipulated altmetrics data. This can be carried out via additional qualitative analyses of altmetrics sources, as Haustein, Peters, Sugimoto, Thelwall, and Larivière (2014) have postulated. The potential manipulation of altmetrics relates also to Campbell's Law, which states, "The more any quantitative social indicator is used for social decision-making, the more subject it will be to corruption pressures and the more apt it will be to distort and corrupt the social processes it is intended to monitor" (Campbell as cited in Sugimoto, 2015, p. 43). This is the case, for example, when HEIs become subject to novel evaluation regimes and make organizational adjustments based on the evaluation criteria. Another example is the "publish or perish" phenomenon, which in some cases may decrease the quality of publications or even result in the fabrication of findings for the sole purpose of meeting set research-output targets. Examples of this issue have been discussed on dedicated blogs, such as Retraction Watch (2017), and at an international biannual conference on research integrity (WCRI, 2017). Additionally, Internet users frequently discuss these false publications on social media, and ironically these discussions subsequently increase altmetrics counts.

As mentioned above, altmetrics are provided by data aggregators (Erdt, Nagarajan, Sin, & Theng, 2016), and the most prominent ones are Altmetric.com, Plum Analytics, Impactstory, and PLoS ALM (Gauch & Blümel, 2016). The San Francisco-based Public Library of Science (PLoS) developed PLoS ALM (or Article-Level Metrics) for its own journals. Impactstory received funding from the U.S. National Science Foundation and the Alfred P. Sloan Foundation (Impactstory, n.d.), and its website allows researchers to showcase their impact in an online profile. Compared to Altmetric.com, Plum Analytics is a secondary data aggregator because its data are collected from secondary sources (Gauch & Blümel, 2016). The Dutch publishing house Elsevier owns the Philadelphia-based Plum Analytics while the London-based Altmetric.com is a portfolio company of Digital Science, belonging to the German Holtzbrinck Publishing Group (Carpenter, 2017; OECD, 2016). The interest in altmetrics shown by such large corporations demonstrates the value of altmetrics, and it indicates which stakeholders might be interested in its generated data.

Marketing materials published by Plum Analytics claim that the company's Dashboard offers several benefits, such as interaction with research users, the potential for new collaborations, and the identification of publication outlets and research funding (Chant, 2016). Still, some studies have suggested that the use of these dashboards might not be widespread, and that most stakeholders in higher education may be unaware of altmetrics (Fraumann, 2017).

Figure 4.1 presents an example of an altmetrics detail page for the publication with the highest score, as tracked by Altmetric.com. The different altmetrics sources, the geographical coverage, and the aggregated altmetric attention score appear in the form of a colorful donut, the colors representing each source for which Altmetric.com aggregated data. The screenshot below (Figure 4.1) shows the number of times Internet users mentioned the article on Facebook, Twitter, Google+, Wikipedia, Q&A sites, video sites, and blogs. Further statistics include how many Mendeley users read the article and how often it appeared on news sites, and in policy documents. The latter is not visible in this screenshot. The headers of these sources can be adapted to the content and context of the article. Users can also sort articles according to the demographics of the users for whom Altmetric.com detected data. The calculation of the score might not be immediately obvious, because mentions on social media weigh differently, as mentioned above.

## Challenges Associated With Altmetrics

Altmetrics address such questions as who mentions the publications of a HEI, a research institute, or a particular scholar, and what are the common interests of these users. These aspects constitute so-called communities of attention (Costas, 2015). Through altmetrics, authors might gain new insights about the users interested in their research outputs. Therefore, they could establish new networks or compare the number of online mentions against similar institutions, because they can identify users of their research, information that would have gone unnoticed without altmetrics. For example, a scholar can find mentions of their research on various sites, such as Twitter, blogs, and news sites culminated in one altmetrics dashboard. Nevertheless, this kind of attention does not necessarily indicate positive attention.

High altmetrics scores based on many interactions, tweets for instance, can also mean that a paper of relatively low quality or one containing a dramatic error has been commented on numerous times, making it an object of humor or even ridicule for many Twitter users (Costas, 2015). One example from Altmetric.com includes an article in which the authors failed to remove insulting comments about the works of other scholars before publication. Another example includes a retracted article in which the first author had suggested a peer reviewer, carried out the peer review on their own, and then sent it to the publisher from a fake e-mail account (Altmetric.com, 2016; Retraction Watch, n.d.). These examples further illustrate that it is essential to know what is behind altmetrics counts and not to simply use them as rankings with which to appraise scholars and their work. Finally, another topical issue is the effect of influential users within online social networks, as this can increase the speed and reach of sharing the news of other users. News sharing networks are playing an increasingly important role in research dissemination (National Academies of Sciences, 2017).

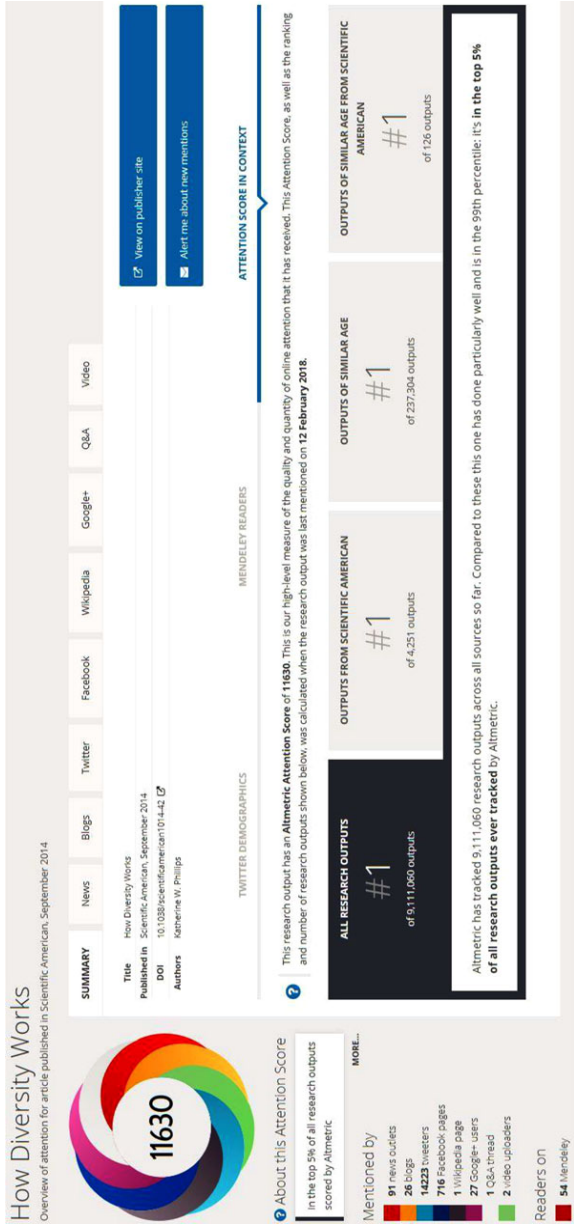

Figure 4.1. Article with the highest altmetric attention score. Adapted from "How Diversity Works" by Katherine W. Phillips, published in the September 2014 issue of *Scientific American* (data compiled by Altmetric.com, as of February 26, 2018)

Scholars and researchers have questioned the reliability of altmetrics; although the understanding of the concept is improving, it is still quite limited. For example, the amount of attention a publication garners on the Internet does not always correlate with its quality (Madjarevic & Davies, 2016), and scholars have raised certain criticisms of altmetrics in this regard (Boon & Foon, 2014). As Haustein et al. (2016) and Holmberg (2014) have suggested, altmetrics data should always be corroborated by qualitative analysis, for instance to identify automated responses (e.g., tweets) generated by bots. The qualitative analysis needs to be performed by comparing various sources manually, although altmetrics data aggregators can also use algorithms to spot suspicious sources.

Scholars have also expressed concerns that altmetrics data are offered mainly by commercial companies, such as Altmetric.com and Plum Analytics (Costas, Zahedi, & Wouters, 2014; Zahedi, Costas, & Wouters, 2014). This is similar to the issue with traditional bibliometrics and proprietary citation databases, such as Scopus and Web of Science (J. Haapamäki, personal communication, October 3, 2016). Moreover, in other research fields, some stakeholders have called for increased university-business relations, whereas others have criticized them. In fact, the research field on altmetrics currently depends primarily on these companies. Relatedly, these companies support the largest annual international meeting, the Altmetrics Conference and Workshop, and they take part in events and discussions with researchers, librarians, and publishers. Current initiatives include also the development of altmetrics dashboards that are based on open software code, independent from companies.

Several studies on altmetrics compare altmetrics data aggregators and examined the differences in their coverage (Jobmann et al., 2014; Zahedi, Fenner, & Costas, 2014). This is connected to the call for altmetrics standards, advocated by many stakeholders, particularly research funders (NISO, 2016). On the basis of a consultation with several stakeholders, a 2016 initiative by the U.S. National Information Standard Organization (NISO) defined certain standards for altmetrics, such as data quality and definitions of key terms. Altmetrics data aggregators were part of this initiative.

Concerning the relation to the academic career model, altmetrics might transform the reward system to a certain extent. For early-career, as well as for senior researchers, it typically takes a relatively long time until journal editors and reviewers approve a manuscript for publication in a journal. In the current academic reward system, these kinds of publications are needed to substantially advance one's career. However, online mentions are within reach of any researcher. Research receiving mentions, shares, downloads, retweets, or comments are relatively faster than citation counts or even a publication in a journal (Bornmann, 2014). Online mentions might increase the motivation to aim for an academic career as they provide early-career researchers a way to develop a stronger presence within their scientific community. At the same time, senior researchers are not disadvantaged if they are inactive on online social networks because altmetrics counts automatically compute even without an online presence. Taking into account the challenges of altmetrics, the inclusion of altmetrics data in assessments of scholarly merits is not currently an option. The validity of the data needs to be enhanced, and the potentially negative effects of introducing a new metric need to be scrutinized (Erdt et al., 2016).

**Ethical Issues Associated With Altmetrics**

Ethical questions concerning altmetrics mostly revolve around the fact that all data are tracked, regardless of whether an online user is aware. This is largely platform dependent; for instance, a Twitter user might expect to be mentioned somewhere else, but Facebook users might be unaware of the extent to which their public profiles and anonymous data about other activities are accessible to external organizations. An example illustrating this issue concerns Mendeley, a reference management software and an online social network created by Elsevier. Mendeley users might be unaware that the owning company, Elsevier, analyzes usage data and that altmetrics data aggregators include it as an altmetrics data source. In the same vein, the Association of Internet Researchers (AoIR) note in its ethical guidelines on the tensions between public and private in the digital age:

> People may operate in public spaces but maintain strong perceptions or expectations of privacy. Or, they may acknowledge that the substance of their communication is public, but that the specific context in which it appears implies restrictions on how that information is—or ought to be—used by other parties. (Markham & Buchanan, 2012, p. 6)

The dilemma of "perceived privacy" is a challenge for other Internet technologies as well, and it is the basis for many conflicts surrounding the inclusion of private data in altmetrics. On the one hand, individual users are not completely visible in large aggregated altmetrics datasets, while, on the other hand, some users might give their consent for data collection while some might not. This depends on individual assumptions and cultural habits, such as the extent to which public online interactions and privacy restrictions are valued. For example, Williams, Burnap, and Sloan (2017) investigated the ethical issues of using Twitter data by employing a large survey of Twitter users. The authors asked survey respondents about the extent to which they would agree that their Twitter data might be useful for publications, such as in research studies. The study findings indicate that most users would not feel comfortable with it, even if using Twitter data does not violate the company's rules. Given this fact, the authors suggest the use of guidelines in asking for informed consent, even when the so-called public social media platforms are part of the data collection.

**Use of Altmetrics**

Several HEIs use altmetrics to promote their research outputs, for example in press releases, including the University of Manchester and Duke University (Madjarevic & Davies, 2016). Furthermore, several online platforms, university library repositories, information systems, and journal websites display altmetrics. Nevertheless, studies have suggested that registered users of these platforms seem to use them irregularly, and they seem largely unaware of the concept of altmetrics (Fraumann, 2017). Several recent studies have focused on a particular system and its altmetrics data (Erdt et al., 2016; Gauch & Blümel, 2016).

What is more, some funders, such as Autism Speaks, the largest international funder for autism research, already connects altmetrics data with their own data about awarded grants to demonstrate the impact of their funded research. The key issue, therefore, has to do with the value research funding organizations and researchers attach to altmetrics counts and rankings. These values are particularly important in research funding, because they might influence funding decisions made by board and committee members. Rankings, according to altmetrics data aggregators, show a simplified output because they aggregate the various counts, such as the number of tweets. Concerning the future promotion of research impact, altmetrics offer a form of measurement that might help answer questions regarding the return on investment for funders by going beyond qualitative reports, such as impact case studies compiled by the funded researchers themselves. A valid approach might be to understand altmetrics not as an auditing tool, but as a way to facilitate the identification of the networks of research users (Robinson-García, van Leeuwen, & Ràfols, 2017).

**The Potential Role of Altmetrics in Institutional Research**

This section provides an overview on the relation between altmetrics and IR. According to the Association for Institutional Research (AIR), the general duties and functions of IR are to "(1) identify information needs; (2) collect, analyze, interpret, and report data and information; (3) plan and evaluate; (4) serve as stewards of data and information; and (5) to educate information producers, users, and consumers." (AIR, 2017) The above-mentioned duties and functions also relate to altmetrics. At its current stage of development, altmetrics are not decision-making tools. Still, there are many categories of work in which IR professionals may consider altmetrics, for example in collaboration with other departments at the same institution, such as libraries and communication departments. Table 4.1 provides an overview of how some IR tasks relate to altmetrics. The list of tasks relies on a national study of IR work tasks that the AIR (2016) published. For the study, the AIR carried out a survey among its members to develop an inventory of tasks performed by senior IR/IE (institutional effectiveness) officers. The final report grouped IR tasks in major categories of work, and the following eleven categories, which represent thirty-two IR tasks that might have a substantive relationship to work on altmetrics as well:

1. Assessment (two tasks);
2. committee work (five tasks);
3. data integrity (seven tasks);
4. educator (one task);
5. management (one task);
6. technology (one task);
7. personal attributes and work (four tasks);
8. planning (two tasks);
9. policies and procedures (one task);
10. reporting (one task); and
11. research (seven tasks).

Table 4.1 includes only the major categories of work that are relevant to altmetrics, hence, some categories are not part of this table and the list above (for further discussion of IR tasks, see AIR, 2016 and Chapter 1 in this volume). IR and altmetrics are primarily data-driven fields that evolve constantly, and the relation between the two might need an update over time.

## Conclusions

This chapter provided an overview of the concept of altmetrics and its relations to IR. Several HEIs in the United States and globally display altmetrics data as part of their demonstration of research impact. Many HEI's libraries and/or IR offices administer this kind of service licensed from altmetrics data aggregators. Due to the persistent challenges related to altmetrics, these services are typically not intended for research evaluation or similar procedures, but rather present an opportunity for universities to showcase public recognition or impact. Consequently, stakeholders are able to explore the mentions of research produced by an HEI or an individual faculty's member via new avenues, such as social media. Altmetrics seem to be on the rise in the online exploration and discussions of scholarly impact.

Table 4.1. Major Categories of Work, IR Tasks, and Their Relations to Altmetrics

| No. | Major Categories of Work | IR Tasks | IR Tasks and Their Relations to Altmetrics |
|---|---|---|---|
| 1 | Assessment | "Assesses the usefulness of services" | Assesses the usefulness of altmetrics data aggregators and altmetrics sources. |
| 2 | Assessment | "Ensures college outreach efforts are assessed" | Compares college outreach efforts with altmetrics data. |
| 3 | Committee work | "Serves on data policy groups at the institution" | Data policy concerning altmetrics. |
| 4 | Committee work | "Serves on institutional committees focused on communicating institutional data" | Communicates institutional data, such as altmetrics counts. |
| 5 | Committee work | "Serves on institutional committees focused on data" | Data, such as altmetrics. |
| 6 | Committee work | "Serves on institution's technology teams" | Technology teams on online sources, such as altmetrics. |
| 7 | Committee work | "Represents the institution on various state/federal data policy groups" | Data policy groups on altmetrics. |
| 8 | Data integrity | "Acts as a consultant to campus depart-ments/colleagues on assessment metrics" | Consults on debates concerning assessment metrics, such as discussions of the use of altmetrics. |
| 9 | Data integrity | "Collaborates with campus colleagues to ensure campus data are integrated" | Collaborates with campus colleagues, such as librarians to ensure data integration with altmetrics and other data points. |
| 10 | Data integrity | "Coordinates institutional data" | Coordinates altmetrics data. |
| 11 | Data integrity | "Gathers data from a variety of sources" | Gathers data from several altmetrics sources. |
| 12 | Data integrity | "Identifies information resources" | Identifies information resources through altmetrics. |
| 13 | Data integrity | "Meets the evolving data/information needs" | Considers altmetrics data as evolving data/information needs. |

(*Continued*)

**Table 4.1. Continued**

| No. | Major Categories of Work | IR Tasks | IR Tasks and Their Relations to Altmetrics |
|---|---|---|---|
| 14 | Data integrity | "Provides information for various institutional reports" | Reports on altmetrics data. |
| 15 | Educator | "Educates campus colleagues on effective use of institutional data" | Educates campus colleagues on effective use of altmetrics data. |
| 16 | Management | "Facilitates cooperation between IR office and other campus offices" | Facilitates cooperation on altmetrics between IR office and other campus offices, such as the library and communications department. |
| 17 | Technology | "Writes queries to pull data from database" | Writes queries to pull data, for example, from altmetrics databases. |
| 18 | Personal attributes and work | "Engages with data partners" | Engages with altmetrics data aggregators. |
| 19 | Personal attributes and work | "Maintains current knowledge in the field of institutional research" | Maintains current knowledge about altmetrics and how they relate to IR. |
| 20 | Personal attributes and work | "Understands higher education and the underlying dynamics and forces that affect it" | Understands the concept of altmetrics in a digital era. |
| 21 | Personal attributes and work | "Reviews news releases" | Identifies news releases about the institution's research through altmetrics. |
| 22 | Planning | "Analyzes reports prepared by state and federal agencies to evaluate their impact on the institution's goals and objectives" | Analyzes reports related to altmetrics and their impact on the institution's goals and objectives. |
| 23 | Planning | "Participates in environmental scanning projects" | Identifies emerging topics through altmetrics. |
| 24 | Policies and procedures | "Interprets college higher education policies" | Altmetrics also relate to college higher education policies. |
| 25 | Reporting | "Creates institution's web-based Fact Book" | The institution's Web-based Fact Book may include information on altmetrics. |

Table 4.1. Continued

| No. | Major Categories of Work | IR Tasks | IR Tasks and Their Relations to Altmetrics |
|---|---|---|---|
| 26 | Research | "Evaluates alternate methods of data collection" | Alternate methods of data collection relate to the concept of altmetrics. |
| 27 | Research | "Collaborates with senior leadership to identify research to support institutional priorities" | Identifies research to support institutional priorities through altmetrics. |
| 28 | Research | "Identifies research projects" | Identifies related research projects through altmetrics. |
| 29 | Research | "Understands data mining" | Understands how altmetrics sources are mined. |
| 30 | Research | "Performs public policy research" | Altmetrics provide an overview if research outcomes appear in policy documents. |
| 31 | Research | "Determines procedures for reporting research" | Altmetrics may support the development of strategies for reporting research to the wider public. |
| 32 | Research | "Understands what statistical analysis can bring to light and what it can mask" | Understands how to perform and interpret statistical analysis regarding altmetrics. |

Several challenges concerning altmetrics need addressing, such as the validity of altmetrics sources, the value of online user engagement with scholarly outcomes, and ethical concerns on the retrieval of altmetrics data. When it comes to institutional platforms' user statistics, the usage of altmetrics is still rather low and knowledge among most stakeholders in higher education is low, as well. However, the years ahead will likely witness an increase in the awareness and use of altmetrics in scholarly communications and beyond and understanding how it works, strengths, and shortcomings will individuals and organizations enable to use altmetrics responsibility and effectively.

*Grischa Fraumann is a research assistant on altmetrics at the German National Library of Science and Technology in Hanover, Germany.*